\documentclass{article}
\usepackage{latexsym}
\usepackage{amsmath,amssymb}
\usepackage{float}
\usepackage[dvipdfmx]{graphicx}%

\newcommand{\Slash}[1]{{\ooalign{\hfil/\hfil\crcr$#1$}}}

\begin{document}

\pagestyle{plain} 
\setcounter{page}{1}
\setlength{\textheight}{700pt}
\setlength{\topmargin}{-40pt}
\setlength{\headheight}{0pt}
\setlength{\marginparwidth}{-10pt}
\setlength{\textwidth}{20cm}

\title{Remark on Structure of Expectation Values of Flavor-Lepton Numbers with respect to Neutrino-Source Hadron States:\\
 Deviation from Fermi's Golden Relation}
\author{Kanji Fujii$^1$ and Norihito Toyota$^2$  \and  \hspace{3mm} 1)Department of Physics, Faculty of Science, Hokkaido University,\\ Sapporo 060-0810, Japan  
\and 2)Faculty of Business Administration and Information Science, \\ \hspace{6mm} Hokkaido Information University, Ebetsu, Nisinopporo 59-2, Japan\\
fujii@particle.sci.hokudai.ac.jp,  toyota@do-johodai.ac.jp }
\date{}
\maketitle

\begin{abstract}
In our preceeding reports, we have pointed out that a unified description of weak decays accompanying neutrinos and the oscillation process is obtained on the basis of the expectation values of flavor-neutrino numbers with respect to the neutrino-source hadron state. In the present report, we investigate the effect on the expectation values due to the deviation from Fermi's golden relation, and give concrete features of these deviations in the case of $\pi^+$ and $K^+$-decays under the simple situation with the $3$-momentum $\vec{p_A}=0$ for $A=\pi^+$, $K^+$. 
  \end{abstract}

\section{ Introduction }
 \subsection{Purpose} 
 
 It has been pointed out in the preceeding papers Ref.\cite{F1,F22,F23} that a unified description of the weak decays accompanied by  neutrinos as well as the neutrino-oscillation process is obtained when we consider the expectation values of the flavor-lepton numbers with respect to  neutrino-source hadron state $A$. 
 (In Ref.\cite{F23} and also in the following we consider mainly the cases $A=\pi^+$ and $K^+$.)
 
 From the expectation-value approach, we obtain the quantities which are natutally interpreted as the decay probabilities  where the neutrino-mixing exists, and also the quantities as the neutrino-oscillation formulas. 
 
 Being motivated by Ref.\cite{Ishi}, we have considered in Ref.\cite{F23} the ration $X_1/X_2$ of the two relevant quantities $X_1$ and $X_2$ in order to see the situation how Fermi's golden relation holds. 
 In the following, after remembering the structures of the ration $X_1/X_2$ given in Ref. \cite{F23}, we examine the difference $X_1-X_2$ between those two relevant quantities; the difference $X_1-X_2$ corresponds to the difference examined in Ref.\cite{Ishi}, and  
the these two kinds of quantities,  $X_1/X_2$ and $X_1-X_2$ are simply related with each other through the fundamental arithmetic operations. This situation is to be confirmed through numerical calculations. 

 \subsection{Definitions of relevant quantities} 
In this subsection we summarize the definitions of relevant quantities, which are used in our preceding work Ref.\cite{F23}. 

The lowest-order contribution to the expectation values of charged-lepton number $N_{\ell \sigma}(x^0)$ mainly comes from the expectation value   

 \begin{equation}
\langle N_{\ell_\sigma},A^+;x^0,x_I^0\rangle_I = \langle A^+(x^0)| \int_{x_I^0}^{x^0} d^4y \int_{x_I^0}^{x^0} d^4z  H_{int}(z)N_{\ell_\sigma}(x^0)H_{int}(y)|A^+(x_I^0)\rangle. \label{eq:1.1a}
\end{equation}
 The dominant contribution comes from the diagram shown by Fig.1. 
\begin{figure}[!h]
\centering
\includegraphics[bb=0 0 360 150,clip,scale=0.75]{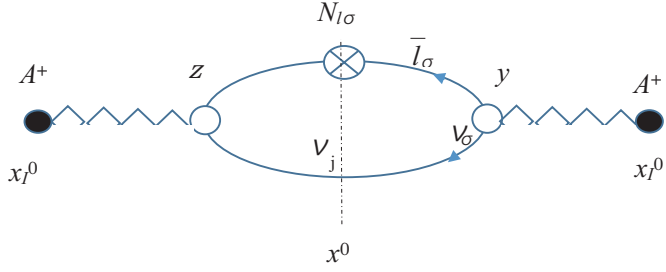}
\caption{Dominant contribution to $\langle N_{\ell \sigma}, A^+(p);x^0,x_I^0\rangle $. 
This diagram is the same as Fig.2 in Ref.\cite{F23}.     }
\label{fig:Fig2}
\end{figure}

 We write this dominant contribution as  $-\langle \bar{n}_{\ell_\sigma},A^+(p);x^0-x_I^0=T \rangle $.  
 Concretely we obtain the the absolute value of the expectation value of $\bar{\ell}_\sigma$-number 
 \begin{eqnarray}
\langle \bar{n}_{\ell \sigma}, A^+(p);x^0-x_I^0=T\rangle \cong \Bigl[\frac{G_Ff_A}{\sqrt{2}}\Bigr]^2  \frac{1}{2E_A(p)V} \sum_{\vec{q}}\sum_{\vec{k}} \delta(\vec{p},\vec{k}+\vec{q})   \nonumber\\
\times \sum_{j} Z_{\sigma j}^{\frac{1}{2}}  Z_{\sigma j}^{\frac{1}{2}*}  R(jjk,\sigma q,p) \Bigl[ \frac{ \sin (T  ( E_A(p)-E_\sigma(q)-\omega_j(k))/2) }{  ( E_A(p)-E_\sigma(q)-\omega_j)/2 }\Bigr]^2,   \label{eq:1.2c}%
\end{eqnarray}
where  

  \begin{equation*}
  R(jjk,\sigma q,p)= \frac{2}{ \omega_j(k) E_\sigma(q)} \big[ m_A^2(k^{(j)}\cdot q)+2 (q\cdot p)(k^{(j)} \cdot p) \big];  \label{eq:1.2}%
\end{equation*}
with the notation $  (k^{(j)}\cdot q) = \vec{k}\vec{q}- \omega_j(k) E_\sigma(q) $ and the neutrino-mixing matrix $\big[ Z_{\sigma j}^{\frac{1}{2}} \big]$, appearing in the relation between the flavor- and mass-neutrino fields, $i.e.$ $\nu_\rho (x)= \sum_{j=1}^3  Z_{\rho j}^{\frac{1}{2}} \nu_j(x)$, for $\rho= e,\mu, \tau$. (See Eqs. (27) $\sim$ (29) in Ref.\cite{F23}.)

We can define the decay amplitude due to the charged-current $H_{int}(z)$ used for deriving  (\ref{eq:1.2c});
\begin{eqnarray}
&&\mathcal{A} (A^+ (p) \rightarrow \bar{\ell}_\sigma (q,s)+\nu_j(k,r);x^0-x^0_I=T) \nonumber \\
&&:=\langle \bar{\ell}_\sigma (q,s)+\nu_j(k,r);x^0 \big|-i \int d\vec{z} \int_{x_I^0}^{x^0}dz^0  H_{int}(z)\big|A^+ (p);x_I^0\rangle. \label{eq:1.3}
\end{eqnarray}
(Here, $s$ and $r$ represent the helicity.) 
As pointed out in Ref.\cite{F23}, we obtain 
 \begin{equation}
\sum_j \sum_{\vec{q},s} \sum_{\vec{k},r} \Bigl| \mathcal{A} (A^+ (p) \rightarrow \bar{\ell}_\sigma (q,s)+\nu_j(k,r);T) \Big|^2 = \langle \bar{n}_{\ell\sigma}, A^+ (p),x^0-x^0_I=T \rangle. \label{eq:1.4}%
\end{equation}
The decay probability $P(A^+(p)\rightarrow \bar{\ell_\sigma(q)}+\nu_j(k)$ for energetically allowed $\ell_\sigma$ per unit time is  
\begin{eqnarray}
P\bigl(A^+ (p) \rightarrow \bar{\ell}_\sigma(q)+\nu_j(k)\bigr) := \sum_{\vec{q},s}\sum_{\vec{k},r}\Bigl[ \big| \mathcal{A}( P(A^+ (p) \rightarrow \bar{\ell}_\sigma(q)+\nu_j(k));T\big|^2 /T \Bigr]_{T\rightarrow large} \nonumber \\
=\Bigl[\frac{G_F f_A}{\sqrt{2}} \Bigr]^2 |Z_{\sigma j}^{1/2}|^2 \int \frac{d \vec{q}}{(2\pi)^3}\cdot \frac{2\pi\delta\bigl(E_A(p) -E_{\sigma}(q)-\omega_j(k)\bigr)}{2E_A(p)}   R(jjk,\sigma q,p) \Bigl|_{\vec{p}=\vec{q}+\vec{k}}.\label{eq:1.5}
\end{eqnarray}

In the above derivation, we have used the relation for a large $T$, 
\begin{equation}
 \Bigl[ \frac{ \sin (T  ( E_A(p)-E_\sigma(q)-\omega_j(k))/2) }{  ( E_A(p)-E_\sigma(q)-\omega_j(k))/2 }\Bigr]^2 \cong 2\pi \delta \bigl(     E_A(p)-E_\sigma(q)-\omega_j(k)\bigr)  T, \label{eq:1.6}%
\end{equation}
called Fermi's golden relation. (See Ref.\cite{comm}.) 
From Eqs.   (\ref{eq:1.4}) and  (\ref{eq:1.5}), we obtain the relation in the case of energetically allowed $\bar{\ell}_\sigma$ 
\begin{equation}
\Big[\langle \bar{n}_{\ell \sigma}, A^+(p);x^0-x_I^0=T\rangle/T \Bigr] _{T\rightarrow large}= \sum_j P\bigl(A^+ (p) \rightarrow \bar{\ell}_\sigma(q)+\nu_j(k)\bigr)  \label{eq:1.7}%
\end{equation}

Under the condition $m_A>m_\sigma +m_j$ where $m_\sigma$ is the mass of $\ell_\sigma$, we obtain 
\begin{eqnarray}
P\bigl(A^+ (\vec{p}=0) \rightarrow \bar{\ell}_\sigma (q)+\nu_j (k) \bigr)
=\frac{(G_F f_A)^2}{8\pi}  |Z_{\sigma j}^{1/2}  |^{2} m_A m_\sigma^2 \Bigl( 1+\frac{m_j^2}{m_\sigma^2}-\frac{(m_\sigma^2-m_j^2)^2}{m_A^2 m_\sigma^2} \Bigr)\nonumber \\
\times \sqrt{ \bigl\{1-\frac{(m_\sigma+m_j)^2}{m_A^2}\bigr\} \bigl\{1-\frac{(m_\sigma-m_j)^2}{m_A^2}\bigr\} }; \label{eq:1.8} 
\end{eqnarray}
due to $m_j/m_\sigma<<1$
\begin{equation*}
R.H.S. of Eq. (\ref{eq:1.8}) \simeq \frac{(G_F f_A)^2}{8\pi}  |Z_{\sigma j}^{1/2}  |^{2} m_A m_\sigma^2 \Bigl( 1-\frac{m_\sigma^2}{m_A^2}\Bigr) ^2. \label{eq:1.8c} 
\end{equation*}
Thus we obtain

 \begin{eqnarray}
\sum_{j} P\bigl(A^+ (\vec{p}=0) \rightarrow \bar{\ell}_\sigma (q)+\nu_j (k)\bigr) &\cong& 
\frac{(G_F f_A)^2}{8\pi}  m_A m_\sigma^2 \Bigl( 1-\frac{m_\sigma^2}{m_A^2}\Bigr)^2 \nonumber \\
&:=& P_0\bigl(A^+ (\vec{p}=0) \rightarrow \bar{\ell}_\sigma+\nu (mass=0) \bigr),  \label{eq:1.9}%
\end{eqnarray}
which describes well the experiment. 
Therefore, L.H.S of (\ref{eq:1.7}) gives the probability (per unit time) of leptonic two-body decay of the mother $A^+$-meson.  

In our preceding paper Ref.\cite{F23}, we have defined the ratio 
\begin{equation}
R\bigl(A^+ (\vec{p}=0), \bar{\ell}_\sigma;T\bigr) :=  \frac{ \langle n_{\ell \sigma}, A^+ (p=0);T\rangle\;\mbox{ with all }m_j's=0}{ P_0\bigl(A^+ (\vec{p}=0) \rightarrow \bar{\ell}_\sigma+\nu(mass=0)\bigr)}. \label{eq:1.10}
\end{equation}
 For convenience, we use hereafter the parameter $L=T\cdot c$ instead of $T$. 
 We see from (\ref{eq:1.7}) and (\ref{eq:1.9}) that the deviation of the ratio $R\bigl(A^+ (\vec{p}=0),\mu^+;L)/L  $ 
from 1 gives us a certain information on the deviation from Fermi's golden relation. 
 In Ref.\cite{F23}, the graphs of $R(A^+(\vec{p}=0),\mu^+;L)/L$ have been given as functions of $L/\lambda_A$ for $A=\pi$ and $K$ cases; here $\lambda_A=\hbar/ (m_A c)$ is Compton wave length with the magnitude $\sim 10^{-15}$m. 
 Those graphs, Fig.3 and Fig.4 in Ref.\cite{F23}, teach us that the ratios $R\bigl(A^+,\mu^+;L)/L $ for $A=\pi$ and $K$  
 approach rapidly to 1 for growing $L$ even from a small $L$-value such as $L/\lambda_A \sim $O($10^2$). 
 In Sec.2, we give again those graphs with the aim of seeing more details concerning the deviation from the golden relation. 
 
 Another quantity leading to a measure of the deviation from the golden relation is the difference
 \begin{equation}
R\bigl(A^+ (\vec{p}=0),\mu^+;L \bigr)/\lambda_A-L/\lambda_A.  \label{eq:1.11}
\end{equation}
 Such a quantity as  (\ref{eq:1.11})  corresponds to the quantity investigated by Ishikawa and Tobita in Ref.\cite{Ishi}.
 
 In the following, after remembering the structures of  $R\bigl(A^+ (\vec{p}=0,\mu^+;L)/L  $ investigated in Ref. \cite{F23}, 
we consider the structure of the difference (\ref{eq:1.11}).  

\section{Form of the ratio  $R\bigl(A^+ (\vec{p}=0),\mu^+;L)/L  $} 
The ratio (\ref{eq:1.10}), which is rewriten in a convenient form for examining its structure, is given as
\begin{equation}
 \frac{R\bigl(A^+ (\vec{p}=0),\bar{\ell}_\sigma;T\bigr)}{L} =  \frac{2/\pi}{a_\sigma^2(1-a_\sigma^2)^2} I(A^+(\vec{p}=0), \sigma;L), \label{eq:1.12}%
\end{equation} 
 where
 \begin{equation}
 I(A^+(\vec{p}=0), \sigma;L) = \frac{\lambda_A}{L}   \int_0^\infty db \cdot b^2  
 \bigl( 1-\frac{b}{\sqrt{a_\sigma^2 +b^2}} \bigr) \Bigl[ \frac{\sin\bigl( \frac{L}{2\lambda_A}( 1-\sqrt{a_\sigma^2 +b^2}-b)\bigr)}{( 1-\sqrt{a_\sigma^2 +b^2}-b)/2} \Bigr]^2,\label{eq:1.13}%
 \end{equation}
with 
   \begin{center}
$a_\sigma=\frac{m_\sigma}{m_A}$,      $b=\frac{|\vec{k}|}{m_A}=\frac{k}{m_A}$ and also $\frac{Tm_A c^2}{\hbar}=Tc\cdot \frac{m_A c}{\hbar}=\frac{L}{\lambda_A}.$  
\end{center}
(Eqs. (\ref{eq:1.12}) and (\ref{eq:1.13}) are equal to Eqs. (43) and (44) in Ref. \cite{F23}.)

The dependence of the ratio (\ref{eq:1.12}) on $L/\lambda_A$ are seen from Fig.2 and Fig.3 in the case of $\bar{\ell}_\sigma=\mu^+$. 
When the golden relation is applied to R.H.S of  (\ref{eq:1.13}), R.H.S of  (\ref{eq:1.12}) goes to 1. 
Thus, from the graphs in Fig.2 and Fig.3, we see that the golden relation appropriately holds for a not-so-large value of $L/\lambda_A$. 
\begin{figure}[t]
\centering
\includegraphics[bb=80 550 550 750,clip,scale=0.80]{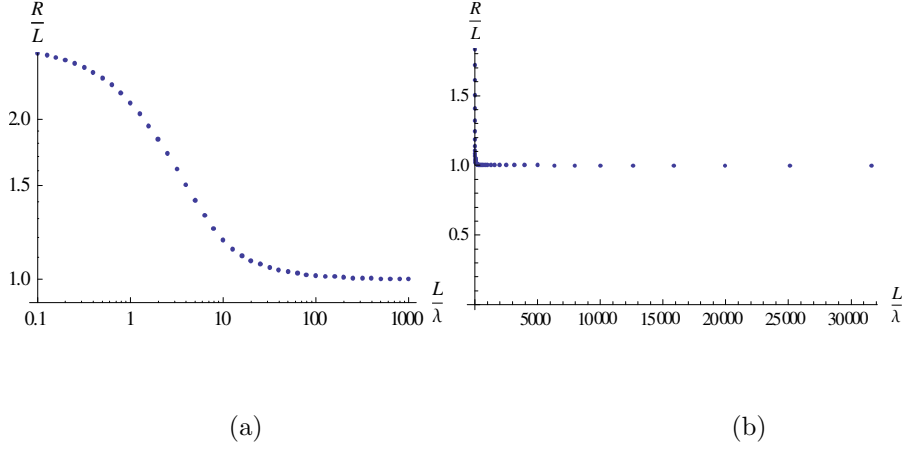}\\
(a) \hspace{60mm} (b)
\caption{$R(\pi^+(\vec{p}=0), \mu^+;L)/L $.  (a) microscopic-$L$ case,  $L/\lambda =0.1\sim 10^3$, 
     (b) intermediate-$L$ case,  $L/\lambda\leq 3\times 10^4$. 
These graphs correspond to Fig.3 (a) and (b) in Ref.\cite{F23}.  
 $\lambda_\pi= 
\frac{197.3 \mbox{Mev}\cdot 10^{-15} \mbox{m}}{139.6\mbox{Mev}},\;\;
a_\mu=\frac{105.7 \mbox{Mev}}{139.6\mbox{Mev}}. 
$
}
\label{fig:Fig3a}
\end{figure}


\begin{figure}[t]
\centering
\includegraphics[bb=0 0 480 170,clip,scale=0.8]{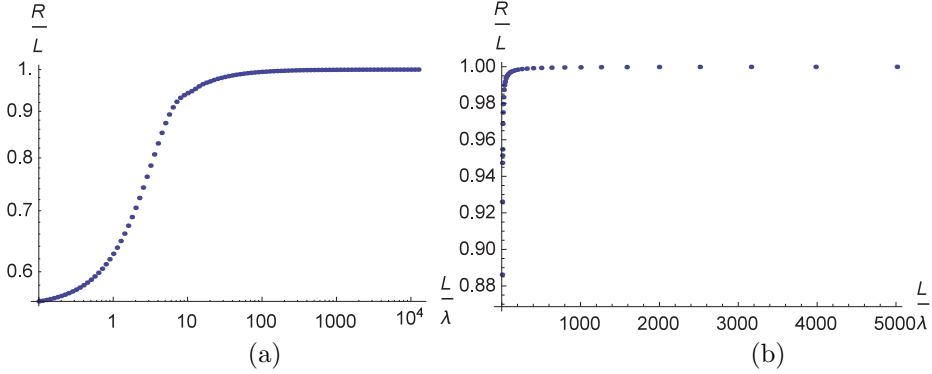}\\
(a) \hspace{60mm} (b)
\caption{$R(K^+(\vec{p}=0), \mu^+;L) /L, $ (a) $L/\lambda\leq 3\times 10^3$,  (b)intermediate-$L$ case.  Fig.3.(a) corresponds to Fig.4 in Ref.\cite{F23}. 
 $\lambda_K= 
\frac{197.3 \mbox{Mev}\cdot 10^{-15} \mbox{m}}{493.7\mbox{Mev}},\;\;
a_\mu=\frac{105.7 \mbox{Mev}}{493.7\mbox{Mev}}. 
$   }
\label{fig:Fig4}
\end{figure}
\section{Structure of the difference $\bigl( R(A^+(\vec{p}=0),\mu^+;L)-L\bigr)/\lambda_A$ }
In this section we examine the difference  $\bigl( R(A^+(\vec{p}=0),\mu^+;L)-L\bigr)/\lambda_A$, for $\lambda_A/L<<1$, 
which gives another possible measure of the deviation from the golden relation, as mentioned in Sec.1-2. 
Hereafter we use for simplicity the notation $R(A,\mu;L)$ instead of $R(A^+(\vec{p}=0), \mu^+;L) $. 

As to the behavior of this difference under $\lambda_A/L\rightarrow 0$, there are three possible cases corresponding to the 
degree how $R(A,\mu;L)/L -1$ tends to 0;\\
\begin{eqnarray}
&\mbox{the difference }&=(\frac{1}{L} R(A,\mu;L)-1)\cdot \frac{L}{\lambda_A} \nonumber \\
&\overrightarrow{\lambda_A/L\rightarrow 0} & \begin{cases}
 0 & if \;\frac{1}{L} R(A,\mu;L)-1 \propto (\frac{\lambda_A}{L})^a  \; with \; a>1, \label{eq:1.14} \\
 const(\neq0) & if   \;\frac{1}{L} R(A,\mu;L)-1 \propto \frac{\lambda_A}{L}, \label{eq:1.15} \\
 \infty &if \;\frac{1}{L} R(A,\mu;L)-1 \propto (\frac{\lambda_A}{L})^a \; with \; 0<a<1.\label{eq:1.16}%
 \end{cases}
 \end{eqnarray}

Concrete numerical calculations of $R(A,\mu;L)/\lambda_A -L/\lambda_A$ for $A=\pi^+$ and $K^+$, shown by Fig.4, lead to the result consistent with the middle case of (\ref{eq:1.14}); i.e. 
\begin{equation}
\frac{1}{\lambda_A} R(A,\mu;L)-\frac{L}{\lambda_A} \simeq \begin{cases}
1.702 & for\; A=\pi^+, \\
-0.583 & for\; A=K^+\label{eq:1.17}%
\end{cases}
\end{equation}
in the range $10^3 \lesssim L/\lambda_A$ $(\lesssim 3\times 10^4)$. 
It seems worthy for us to notice that 
the distributive law $\bigl( R(A,\mu;L)/L-1\bigr)\times \frac{L}{\lambda_A}= R(A,\mu;L)/\lambda_A- L/\lambda_A$ is confirmed numerically to hold well. 

Thus we may write approximately, in the region $L/\lambda_A>>1$, 
\begin{equation}
\frac{1}{L} R(A,\mu;L)-1=J_1(A,\mu)\frac{\lambda_A}{L} + \mbox{terms with the order } \Bigl(\frac{\lambda_A}{L}\Bigl)^n,
\; n \geq 2,\label{eq:1.18}%
\end{equation}
  where 
\begin{equation}
J_1(A,\mu) \simeq 1.702 \mbox{ and } J_1(A,K) \simeq -0.583. \label{eq:1.18c}%
\end{equation}
It will be neaningful to note that the deviations of the ratio $R(A,\mu;L)/L$ from 1 for not-so-large $L/\lambda_A$, 
which are seen from Fig.2(a) and Fig.3(a), are consistent with (\ref{eq:1.18}) and (\ref{eq:1.18c}), because 

$[ Case \; A=\pi ] $ Fig.2(a) leads to 
\begin{equation}
1.2 \simeq \frac{R(\pi,\mu;L)}{L} |_{\frac{L}{\lambda} \simeq 10}  \simeq 1+\frac{J_1(\pi,\mu)}{10}; \label{eq:1.19}%
\end{equation}

$ [ Case \; B=K ]$ Fig.3(a) leads to 
\begin{equation}
0.95\simeq \frac{R(K,\mu;L)}{L}|_{\frac{L}{\lambda} \simeq 10} \simeq 1+\frac{J_1(K,\mu)}{10}.\label{eq:1.20}%
\end{equation}
\begin{figure}[t]
\centering
\includegraphics[bb=90 350 680 550,clip,scale=0.99]{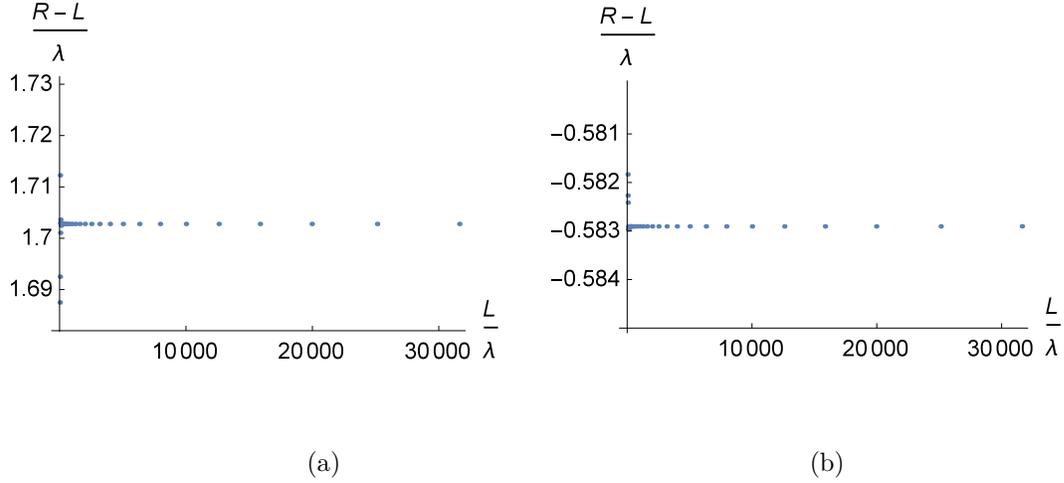}\\
(a) \hspace{60mm} (b)
\caption{Dependence of $\{ R(A, \mu;L)-L \} /\lambda_A (\equiv \tilde{J}(A,\mu;L) )$; (a) the case of $A=\pi$, (b)the case of $A=K$.    }
\label{fig:Fig4}
\end{figure}
 \section{Additional remark on  $ R(A,\mu;L)/\lambda_A - L/\lambda_A$ }
 For simplicity, we use the notation 
 \begin{equation}
  \tilde{J} (A,\mu;L)=R(A,\mu;L)/\lambda_A - L/\lambda_A.\label{eq:1.21}%
\end{equation}
 From  (\ref{eq:1.18}), we have
  \begin{equation}
\tilde{J} (A,\mu;L)=J_1(A,\mu)+\mbox{ terms with the order} \;\; (\frac{\lambda_A}{L})^{n-1}, \;\; n \geq 2.\label{eq:1.22}%
\end{equation}
 With the aim of seeing why $\tilde{J} (A,\mu;L) $ has this structure, we consider the following form of  $\tilde{J} (A,\mu;L) $
  obtained from (\ref{eq:1.12}) ;
  \begin{equation}
  \tilde{J}(A,\mu;L)=\Bigl( \frac{2/\pi}{a^2_\mu (1-a_\mu^2)^2}I(A,\mu;L)-1 \Bigr)\frac{L}{\lambda_A}\label{eq:1.23}%
  \end{equation}
At present for convenience, we employ the notation
  \begin{equation}
  F(A,\mu;L) :=I(A,\mu;L)L/\lambda_A;\label{eq:1.24}%
\end{equation}
then,
  \begin{equation}
  \tilde{J} (A,\mu;L)=  \frac{2/\pi}{a^2_\mu (1-a_\mu^2)^2}F(A,\mu;L) -\frac{L}{\lambda_A}\label{eq:1.25}%
\end{equation}
 By using the concrete form of $I(A,\mu;L)$, $i.e.$ Eq. (\ref{eq:1.13}), we have
 
 \begin{eqnarray}
 \frac{dF(A,\mu;L)}{dL}&=& \frac{2}{\lambda_A}  \int_0^\infty db \cdot b^2  
 \bigl( 1-\frac{b}{\sqrt{a_\sigma^2 +b^2}} \bigr)  \nonumber \\
&\times &\frac{ \sin\bigl( \frac{L}{2\lambda_A}( 1-\sqrt{a_\sigma^2 +b^2}-b)\cos\bigl( \frac{L}{2\lambda_A}( 1-\sqrt{a_\sigma^2 +b^2}-b)\bigr)}{( 1-\sqrt{a_\sigma^2 +b^2}-b)/2}; \label{eq:1.26}%
 \end{eqnarray}
 therefore,
\begin{equation}
\lambda_A  \frac{d\tilde{J}(A,\mu;L)}{dL}= \frac{2/\pi}{a^2_\mu (1-a_\mu^2)^2}\cdot \lambda_A  \frac{dF(A,\mu;L)}{dL}-1\label{eq:1.27}%
\end{equation}
with
 \begin{equation}
\lambda_A  \frac{dF(A,\mu;L)}{dL}= \int_0^\infty db \cdot b^2  
 \bigl( 1-\frac{b}{\sqrt{a_\sigma^2 +b^2}} \bigr)  
\frac{ 2\sin\bigl( \frac{L}{\lambda_A}( 1-\sqrt{a_\sigma^2 +b^2}-b)}{ 1-\sqrt{a_\sigma^2 +b^2}-b} .\label{eq:1.28}%
 \end{equation}
 From Eq. (\ref{eq:1.18}) and Fig.4, $\tilde{J}(A,\mu;L) $ is seen to be nearly $L$-independent and  
\begin{equation} \tilde{J}(A,\mu;L) \simeq J_1(A,\mu) \mbox{ for } 10^3 \lesssim L/\lambda_A. \label{eq:1.29}\end{equation} 
The consistence of our calculation is seen, through numerical evaluation of the integral (\ref{eq:1.28}), by confirming 
 \begin{equation}
 |R.H.S.\; of\;  (\ref{eq:1.27})| \ll 1.\label{eq:1.30}
 \end{equation}
 
 It seems meaningful to note the Dirac delta function is expressed as
 \begin{equation}
 \delta(x)= \lim_{g\rightarrow\infty} \frac{\sin(gx)}{\pi x}.
 \end{equation}
 (See Ref.\cite{comm}.) By applying this relation to R.H.S. of Eq. (\ref{eq:1.28}), 
 we obtain, as expected,
 \begin{equation}
 \lambda_A \frac{dF(A,\mu;L)}{dL} \; \xrightarrow{large\; L/\lambda_A }  \frac{\pi}{2} a^2_{\mu} (1-a^2_\mu)^2.
 \end{equation}
This shows the numerical calculations leading to (\ref{eq:1.30}) 
 have been correctly performed.  

In the considerations described in this report, we have investigated the structures of the ratio $R(A(\vec{p}),\mu;L)/L$ and the difference 
$\bigl( R(A(\vec{p}),\mu;L) -L\bigr)/\lambda_A$ in the case of $\vec{p}=0$ for simplicity. 
These quantities have characteristic features related with the deviation from Fermi's golden relation. 
We hope those features are confirmed experimentally.


\begin{thebibliography}{10}
\bibitem{F1}
K.Fujii and T.Shimomura, Prog.Theor. Phys. \underline{112},901(2004);arXiv:hep-ph/0406079[Search in SPIRE].

\bibitem{F22} 
K.Fujii and T.Shimomura, Phys. of Atomic Nuclei \underline{69}(2006),1353(2006).

\bibitem{F23} K.Fujii and N. Toyota, Prog. Theor. Exp. Phys., \underline{2015}, 023301. DOI:10.1093/ptep/ptu178.











\bibitem{Ishi}
K.Ishikawa and Y. Tobita, Prog.Theor. Exp. Phys. \underline{2013}, 073B02(2013). 


\bibitem{comm} E.g. L.Schiff, "Quantum Mechanics", MaGRAW-Hill, NewYork, 1955
\end{thebibliography}
\end{document}